\documentstyle[eqsecnum,aps]{revtex}

\begin{document}
\draft
\title{Nonequilibrium perturbation theory for spin-$\frac{1}{2}$ fields}
\author{I. D. Lawrie and D. B. McKernan\thanks{Present address:  Department of
Experimental Physics, University College, Dublin, Republic of Ireland.}}
\address{Department of Physics and Astronomy, The University of Leeds,
Leeds LS2 9JT, England}
\date{July 12, 2000}
\maketitle
\begin{abstract}
A partial resummation of perturbation theory is described for field theories containing spin-$\frac{1}{2}$ particles in
states that may be far from thermal equilibrium.  This allows the nonequilibrium state to be characterized in terms of
quasiparticles that approximate its true elementary excitations.  In particular, the quasiparticles have dispersion
relations that differ from those of free particles, finite thermal widths and occupation numbers which, in contrast to
those of standard perturbation theory evolve with the changing nonequilibrium environment.  A description of this
kind is essential for estimating the evolution of the system over extended periods of time.  In contrast to the
corresponding description of scalar particles, the structure of nonequilibrium fermion propagators exhibits features 
which have no counterpart in the equilibrium theory.
\end{abstract}


\section{Introduction}

Many interesting physical problems, arising, for example, in the study of the early universe \cite{abbott86,kolb90,trodden99}
and in relativistic heavy-ion collisions \cite{harris96,rau96} require an understanding of the evolution with time of 
highly-excited states of a quantum field theory.  The properties of high-temperature states in thermal equilibrium have
been studied for a long time and much is known about them (see, for example \cite{lebellac96}).  For the most part,
attempts to study the nonequilibrium properties of systems that evolve with time have been based on the assumption
that this evolution can be adequately represented as a sequence of near-equilibrium states.  While such an assumption
{\it may} be justified in some cases, our understanding of the true nonequilibrium dynamics is at present very incomplete.
One route towards a more complete understanding that has been pursued by several groups is through the study of the
$N\to\infty$ limit of $N$-component scalar field theories (see \cite{cooper97,boyanovsky98,baacke00} and references
cited in these papers).  The major advantage of this limit is that it is a gaussian field theory for which the path integral can
be evaluated exactly, yielding closed-form evolution equations suitable for numerical solution.  However, it is also a theory
devoid of scattering processes, which in general lead to important dissipative effects (and the same is true of the related
Hartree and one-loop approximations).  Moreover, it seems to be extremely difficult to extend such calculations beyond 
leading order in $1/N$, which is essential for approaching any description of more realistic systems.  In this paper, we focus
on the most obvious alternative of extracting as much information as possible from perturbation theory. 

A serious limitation of standard perturbation theory is that, being an expansion about a non-interacting theory, it too is
devoid of scattering at lowest order.  This means, in particular, that the occupation numbers of single-particle modes which
appear in propagators are fixed at some initial values and do not reflect the evolving nonequilibrium state.  Low-order 
calculations therefore become essentially meaningless unless they are restricted to time intervals much shorter than a
typical relaxation time.  To improve this situation, one should reformulate perturbation theory so as to describe the
nonequilibrium state in terms of its own quasiparticle excitations.  These excitations have a nonzero thermal width,
which in part also characterizes the rate of relaxation of their occupation numbers in response to a changing
environment.  To put this idea into practice, it is necessary to construct a lowest-order approximation to the
interacting theory in which at least some of the dissipative effects of interactions are resummed.  Methods for
achieving this in the case of both real and complex scalar theories have been described in \cite{lawrie89,lawrie92,lawrie97}
and incorporated in a comprehensive perturbative approach to the nonequilibrium dynamics of phase transitions in
\cite{lawrie99}.  The purpose of this paper is to investigate how the same idea might be implemented for spin-$\frac{1}{2}$
fields.  The nonequilibrium dynamics of spinor fields turns out to be quite complicated.  In contrast to scalar fields, their
propagators appear to have a structure that is not simply a time-dependent generalization of the one that applies in
thermal equilibrium;  it is sufficiently complicated that we have not been able to explore it in full generality.

We begin in Section II by reviewing briefly the resummation of 2-point functions for real scalar fields.  In Section III, we
derive some general properties of the full spinor 2-point functions which serve as a guide to the construction of an 
effective quasiparticle action, under the simplifying assumption that the latter will be CP-invariant.  The quasiparticle
action is constructed in Section IV in terms of several undetermined functions of time and spatial momentum that
characterize the quasiparticle dispersion relation, thermal width and occupation numbers.  These functions appear
in a counterterm which is added to the free part of the action and subtracted from the interaction part, and will
subsequently be determined self-consistently by requiring the counterterm to cancel part of the higher-order
corrections to the self energy.  The real- and imaginary-time quasiparticle propagators corresponding to this effective
action are derived in Sections V and VI respectively and a self-consistent criterion for determining the quasiparticle
parameters is implemented in the context of a simple model in Section VII.  For illustrative purposes, we introduce
supplementary approximations that allow them to be evaluated in closed form.  These approximations correspond
to a weakly interacting system close to equilibrium, and for this special situation we find, reassuringly, that
the evolution of occupation numbers is described by a Boltzmann equation.  Finally, in Section VIII, we summarize
our principal conclusions and comment on their relation to some other approaches to non-equilibrium field theory.
  
\section{Dissipative perturbation theory for scalar fields}
Consider the usual $\lambda\phi^4$ theory, defined by the Lagrangian density
\begin{equation}
{\cal L}=\frac{1}{2}\partial_{\mu}\phi\partial^{\mu}\phi-\frac{1}{2} m^2(t)\phi^2-\frac{1}{4!}\lambda\phi^4\,,
\label{scalarmodel}
\end{equation}
and suppose that an initial state of thermal equilibrium with inverse temperature $\beta$ is set up at time $t=0$.  In
this model, the time-dependent mass $m(t)$, which arises, for example, in the case of a scalar field theory in a
Robertson-Walker spacetime, drives the subsequent state away from equilibrium. Then the closed-time-path formalism 
(developed for general time-dependent situations in \cite{semenoff85}) yields a path integral weighted by the action
\begin{equation}
I_{\rm C}(\phi_1, \phi_2, \phi_3)=\int{\rm d}^3x\left[\int_0^{\infty}{\rm d}t\,{\cal L}(\phi_1)-\int_0^{\infty}{\rm d}t\,{\cal L}(\phi_2)
+i\int_0^{\beta}{\rm d}\tau\,{\cal L}_{\rm E}(\phi_3)\right]\,,
\end{equation}
where the path-integration variables $\phi_1$, $\phi_2$ and $\phi_3$ live on a closed contour  C in the complex time
plane.  This contour runs along the real axis from $t=0$ to a final time $t_f$, returns along the real axis to $t=0$, and
finally descends along the imaginary axis to $t=-i\beta$.  Here, we have taken the limit $t_f\to\infty$.  The Euclidean 
action ${\cal L}_{\rm E}$ (which uses $m(0)$) represents the initial density matrix.  In this
theory, there is a $3\times 3$ matrix of 2-point functions $G_{ab}(x,x')$, with $a, b = 1, 2, 3$, but our attention will
focus mainly on the real-time functions, with $a, b = 1, 2$.  For the real-time part of the action, we wish to
construct a lowest-order version $I_{\rm C 0}(\phi_1,\phi_2)=-\frac{1}{2}\int{\rm d}^4x\,\phi_a{\cal D}_{ab}\phi_b$, where, after a
spatial Fourier transform, the differential operator ${\cal D}$ is
\begin{equation}
{\cal D}_k(t, \partial_t)=\pmatrix{\partial_t^2+k^2+m^2(t)&0\cr \cr0&-[\partial_t^2+k^2+m^2(t)]}-{\cal M}_k(t,\partial_t)\,.
\label{D1}
\end{equation}
The counterterm $\frac{1}{2}\int d^4x\,\phi_a{\cal M}_{ab}\phi_b$ is added to $I_{\rm C 0}$ and subtracted from the interaction
$I_{\rm C\,int}=I_{\rm C}-I_{\rm C 0}$, so as to leave the whole theory unchanged.  A choice of ${\cal M}$
is a choice of the approximate theory about which we perturb and is, of course, equivalent to a choice of ${\cal D}$.
Subject to several constraints (discussed in \cite{lawrie89}, and generalized below for spinors),
the most general choice for ${\cal D}$ is
\begin{equation}
{\cal D}_k(t,\partial_t)=\pmatrix{[\partial_t^2+\beta_k(t)-i\alpha_k(t)]&[\gamma_k(t)\partial_t+\frac{1}{2}\dot{\gamma}_k(t)
+i\alpha_k(t)]\cr\cr[-\gamma_k(t)\partial_t-\frac{1}{2}\dot{\gamma}_k(t)+i\alpha_k(t)]&
[-\partial_t^2-\beta_k(t)-i\alpha_k(t)]}\,,\label{D2}
\end{equation}
where $\alpha_k(t)$, $\beta_k(t)$ and $\gamma_k(t)$ are real functions yet to be determined.  Of course, the
counterterm ${\cal M}$ can be read from (\ref{D1}) and (\ref{D2}).

The $2\times 2$ matrix of quasiparticle propagators $g_k(t,t')$ is the solution (subject to suitable boundary conditions)
of
\begin{equation}
{\cal D}_k(t, \partial_t)g_k(t,t')=g_k(t,t'){\cal D}_k(t', -\overleftarrow{\partial}_{t'})=-i\delta(t-t')\,.
\end{equation}
Suppressing the spatial momentum $k$, this solution can be written in terms of a single
complex function $h(t,t')$ as
\begin{equation}
g_{ab}(t,t')=h_b(t,t')\theta(t-t')+h_a(t',t)\theta(t'-t)\,,
\end{equation}
where $h_1=h$ and $h_2=h^*$.  The function $h$ is
\begin{equation}
h(t,t')=\frac{1}{2}\exp\left(-\frac{1}{2}\int_{t'}^t\gamma(t'')dt''\right)
\left\{[1+N(t')]f^{(+)}(t)f^{(-)}(t')+[-1+N^*(t')]f^{(-)}(t)f^{(+)}(t')\right\}\,,
\end{equation}
with $f^{(\pm)}(t)=[2\Omega(t)]^{-1/2}\exp\left(\mp i\int_0^t\Omega(t'')dt''\right)$.
We see that one of the undetermined functions, $\gamma_k(t)$, can be interpreted as a quasiparticle
width.  The quasiparticle energy $\Omega_k(t)$ is a solution of
\begin{equation}
\frac{1}{2}\frac{\ddot{\Omega}_k(t)}{\Omega_k(t)}-\frac{3}{4}\frac{\dot{\Omega}_k^2(t)}{\Omega_k^2(t)}
+\Omega_k^2(t)=\beta_k(t)-\frac{1}{4}\gamma_k^2(t)\,.
\label{omegaeqn}
\end{equation}
Finally, the function $N_k(t)$, which we hope to interpret in terms of time-dependent occupation numbers, is a
solution of
\begin{equation}
\left[\partial_t+\gamma_k(t)+2i\Omega_k(t)-\frac{\dot{\Omega}_k(t)}{\Omega_k(t)}\right]\left[\partial_t+\gamma_k(t)
\right]N_k(t)=2i\alpha_k(t)\,.
\label{neqn}
\end{equation}

To give substance to this scheme, a prescription is needed for determining the three
functions $\alpha_k(t)$, $\beta_k(t)$ and $\gamma_k(t)$ introduced in (\ref{D2}).  To this end, define the $2\times 2$
self energy matrix $\Sigma_k(t,t')$ by
\begin{equation}
G_k(t,t')=g_k(t,t')+i\int\,{\rm d}t''\,{\rm d}t'''\,g_k(t,t'')\Sigma_k(t'', t''')G_k(t''',t')\,.
\end{equation}
This self energy has contributions from the counterterm ${\cal M}$ and from loop diagrams:
\begin{equation}
\Sigma_k(t,t')={\cal M}_k(t, \partial_t)\delta(t-t') + \Sigma_k^{\rm loop}(t,t')\,.
\end{equation}
The general strategy is to optimize $g_k(t,t')$ as an approximation to the full two-point functions $G_k(t,t')$ by
arranging for ${\cal M}$ to cancel some part of $\Sigma^{\rm loop}$.  Clearly, since $\Sigma^{\rm loop}$ is non-local
in time, only a partial cancellation can be achieved.  Various prescriptions might be possible;  perhaps the most
obvious is the following.  Express $\Sigma_k(t,t')$ in terms of the average time $\bar{t}=\frac{1}{2}(t+t')$ and the difference
$(t-t')$ and Fourier transform on $(t-t')$.  The components of ${\cal M}_k(t,\partial_t)$ contain at most one time
derivative, so the self energy can be decomposed into contributions that are even and odd in the frequency:
\begin{equation}
\Sigma_k(\bar{t},\omega)={\cal M}_k^{(1)}(\bar{t})+{\cal M}_k^{(2)}(\bar{t})\omega
+\Sigma_k^{(1)\ {\rm loop}}(\bar{t},\omega^2)+\Sigma_k^{(2)\ {\rm loop}}(\bar{t},\omega^2)\omega\,.
\end{equation}
Generalized gap equations to be solved for $\alpha_k(t)$, $\beta_k(t)$ and $\gamma_k(t)$ can now be obtained
by requiring
\begin{eqnarray}
{\cal M}_k^{(1)}(\bar{t})&=&-\Sigma_k^{(1)\ {\rm loop}}\left(\bar{t},\Omega_k^2(\bar{t})\right)\\
{\cal M}_k^{(2)}(\bar{t})&=&-\Sigma_k^{(2)\ {\rm loop}}\left(\bar{t},\Omega_k^2(\bar{t})\right)\,,
\end{eqnarray}
which amounts to an on-shell renormalization prescription.

These gap equations provide exact implicit definitions of $\alpha_k(t)$, $\beta_k(t)$ and $\gamma_k(t)$, but they
cannot, of course, be exactly solved.  If the perturbative expansions for $\Sigma_k^{(1)\ {\rm loop}}$ and
$\Sigma_k^{(2)\ {\rm loop}}$ are truncated at some finite order, one obtains concrete expressions for them
in terms of the propagators $g_k(t,t')$.  These truncated gap equations, together with equation (\ref{omegaeqn})
for the quasiparticle energy and (\ref{neqn}) for the function $N_k(t)$ form a closed system that one might try
to solve numerically.  It is to some extent illuminating to establish a connection with kinetic theory through
some further approximations. Suppose that the gap equations are truncated at two-loop order -- the lowest
order that yields a nonzero quasiparticle width $\gamma_k(t)$.  Then, assuming sufficiently weak coupling
and sufficiently slow time evolution, propagators inside the loop diagrams can be approximated by
taking $\int_{t'}^t\Omega_k(t''){\rm d}t'' \approx \Omega_k(\bar{t})(t-t')$ and the limit $\gamma_k(t)\to 0$.  Then,
with quasiparticle occupation numbers $n_k(t)$ defined by
\begin{equation}
N_k(t)=\frac{1+2n_k(t)}{1-i\gamma_k(t)/2\Omega_k(t)}
\end{equation}
a time-derivative expansion of (\ref{neqn}), yields the Boltzmann-like equation
\begin{eqnarray}
\partial_tn_k(t)&\approx& \frac{\lambda^2}{32(2\pi)^5}\int{\rm d}^3k_1{\rm d}^3k_2{\rm d}^3k_3
\frac{\delta\left(\Omega_1+\Omega_2-\Omega_3-\Omega_k\right)
\delta\left({\bf k}_1+{\bf k}_2-{\bf k}_3-{\bf k}\right)}
{\Omega_1\Omega_2\Omega_3\Omega_k}\nonumber\\
&&\ \ \ \ \times\left[n_1n_2(1+n_3)(1+n_k)-(1+n_1)(1+n_2)n_3n_k\right].
\end{eqnarray}
In the following sections, we investigate how this resummation scheme might be extended to spin-$\frac{1}{2}$
fields.

\section{Exact properties of  spinor 2-point functions}
To be concrete, we consider a system defined by the Lagrangian density
\begin{equation}
{\cal L}=\overline{\psi}\left[\gamma^\mu\partial_\mu-m(t)\right]\psi + \Delta{\cal L}
\label{lagrangian}
\end{equation}
where $\Delta{\cal L}$ represents the coupling of the spinor $\psi$ to other fields.
For the purposes of this work, we suppose once more that the system is driven away from
thermal equilibrium by the time-dependent mass $m(t)$ (and possibly by other
time-dependent parameters in $\Delta{\cal L}$), but remains spatially homogeneous.
As in the scalar case, spinor field theory in a flat Robertson-Walker universe can be represented
as a Minkowski-space theory with time-dependent mass.  If the state at an initial
time that we shall call $t=0$ is one of thermal equilibrium, then standard methods
(described, for example, in Ref. \cite{kobes85}) serve to derive the generating
functional
\begin{equation}
Z(\xi,\overline{\xi})=\int\prod_{a=1}^3[d\overline{\psi}_ad\psi_a]
\ \exp[iI_{\rm c}(\overline{\psi},\psi)+i\,\Xi_{\rm c}(\overline{\psi},\psi,\overline{\xi},\xi)]\,,
\end{equation}
which generalizes that described in Section II for a scalar field.  In this case, the
source term is
\begin{equation}
\Xi_{\rm c}(\overline{\psi},\psi,\overline{\xi},\xi)=\int_{0}^{t_f}dt\left[\overline{\xi}_1(t)\psi_1(t)
+\overline{\psi}_1(t)\xi_1(t)+\overline{\xi}_2(t)\psi_2(t)+\overline{\psi}_2(t)\xi_2(t)\right]
+\int_0^\beta d\tau\left[\overline{\xi}_3(\tau)\psi_3(\tau)+\overline{\psi}_3(t)\xi_3(\tau)\right]
\end{equation}
and we do not indicate explicitly the other fields that may appear in $\Delta{\cal L}$.
If the initial state is characterized by a temperature $\beta^{-1}$ and chemical potential
$\mu$, then the path-integration variables at the ends of the closed time path obey the
boundary conditions $\psi_1(0)=-e^{\beta\mu}\psi_3(\beta)$ and $\overline{\psi}_1(0)
=-e^{-\beta\mu}\overline{\psi}_3(\beta)$, which are inherited by the Green's functions.

As before, we are particularly concerned with the real-time 2-point functions
\begin{equation}
{\cal S}^{(ab)}_{\alpha\beta}(\bbox{x},t;\bbox{x}',t')=\left.
\frac{\delta}{\delta\overline{\xi}_{a\alpha}(\bbox{x},t)}\frac{\delta}{\delta\xi_{b\beta}(\bbox{x}',t')}
\ln Z\right\vert_{\overline{\xi}=\xi=0}
\end{equation}
for $a,b=1,2$.  In terms of field operators, they are
\begin{equation}
{\cal S}_{\alpha\beta}^{(ab)}(\bbox{x},t;{\bf x}',t')=\pmatrix{
\langle \bbox{\cal T}[\psi_{\alpha}(\bbox{x},t)\overline{\psi}_{\beta}(\bbox{x}',t')]\rangle_\mu&
-\langle \overline{\psi}_{\beta}(\bbox{x}',t')\psi_{\alpha}(\bbox{x},t)\rangle_\mu\cr
\cr
\langle \psi_{\alpha}(\bbox{x},t)\overline{\psi}_{\beta}(\bbox{x}',t')\rangle_\mu&
\langle \overline{\bbox{\cal T}}[\psi_{\alpha}(\bbox{x},t)\overline{\psi}_{\beta}(\bbox{x}',t')]\rangle_\mu}\,,
\end{equation}
where $\alpha$ and $\beta$ are spinor indices, while $\bbox{\cal T}$ and $\overline{\bbox{\cal T}}$ denote
time ordering and anti-time ordering respectively.  In the presence of a chemical potential, the
expectation values are given by
\begin{equation}
\langle A\rangle_\mu=\frac{{\rm Tr}\left[e^{-\beta(\widehat{H}-\mu \widehat{N})}A\right]}
{{\rm Tr}\left[e^{-\beta(\widehat{H}-\mu \widehat{N})}\right]}\,,
\label{expectation}
\end{equation}
where $\widehat{H}$ is the Hamiltonian at the initial time and $\widehat{N}$ is the particle number. 

We hope to construct a
perturbation theory in which the lowest-order propagators are partially resummed versions of these full
2-point functions, and begin by establishing some properties of the full functions that our approximate
ones ought to share.  Expecting that correlations should decay, very roughly as $e^{-\lambda\vert t-t'\vert}$,
over large time intervals, we write the Wightman function ${\cal S}^>_{\alpha\beta}(\bbox{x},t;\bbox{x}',t';\mu)
=\langle \psi_{\alpha}(\bbox{x},t)\overline{\psi}_{\beta}(\bbox{x}',t')\rangle_\mu$  as
\begin{equation}
{\cal S}^>_{\alpha\beta}(\bbox{x},t;\bbox{x}',t';\mu)
={\cal H}_{\alpha\beta}(\bbox{x},t;\bbox{x}',t')\theta(t-t')+{\cal K}_{\alpha\beta}(\bbox{x},t;\bbox{x}',t')\theta(t'-t)\,.
\end{equation}
Using $\overline{\psi}_\beta=\psi^\dag_\gamma\gamma^0_{\gamma\beta}$, it is easy to see that
\begin{equation}
{\cal S}^{>\dag}(\bbox{x},t;\bbox{x}',t';\mu)=
\gamma^0{\cal S}^{>}(\bbox{x}',t';\bbox{x},t;\mu)\gamma^0
\end{equation}
and hence that
\begin{equation}
{\cal K}(\bbox{x},t;\bbox{x}',t';\mu) = \overline{\cal H}(\bbox{x}',t';\bbox{x},t;\mu)\,,
\end{equation}
where, for any Dirac matrix, we define $\overline{M}=\gamma^0M^\dag\gamma^0$. It would be helpful
if the second Wightman function ${\cal S}^<_{\alpha\beta}(\bbox{x},t;\bbox{x}',t';\mu)
=-\langle \overline{\psi}_{\beta}({\bf x}',t')\psi_{\alpha}({\bf x},t)\rangle_\mu$  could be expressed in
terms of the same matrix ${\cal H}(\bbox{x},t;\bbox{x}',t';\mu)$.  This can, in fact,
be done in a CP-invariant theory.  If the CP transformation of an operator $A$ is implemented
by a unitary operator $U_{\rm CP}$, so that $A^{\rm CP} = U_{\rm CP}^{-1}AU_{\rm CP}$, then
a CP-invariant theory has $\widehat{H}^{\rm CP}=\widehat{H}$ and $\widehat{N}^{\rm CP}=-\widehat{N}$,
and we see from (\ref{expectation}) that
\begin{equation}
\langle A\rangle_\mu=\langle A^{\rm CP}\rangle_{-\mu}\,.
\end{equation}
For a Dirac spinor, we have $\psi^{\rm CP}(\bbox{x},t)=\gamma_0C\overline{\psi}^T(-\bbox{x},t)$,
where $C$ is the charge conjugation matrix and $^T$ indicates the transpose.  It follows from this
that
\begin{eqnarray}
{\cal S}^<_{\alpha\beta}(\bbox{x},t;\bbox{x}',t';\mu)
&=&\left[C^{-1}\gamma_0{\cal S}^>_{\alpha\beta}(-\bbox{x}',t';-\bbox{x},t;-\mu)\gamma^0C\right]^T
\nonumber\\
&=&\widetilde{\overline{\cal H}}(-\bbox{x},t;-\bbox{x}',t';\mu)\theta(t-t')
+\widetilde{\cal H}(-\bbox{x}',t';-\bbox{x},t;\mu)\theta(t'-t)\,,
\label{cprelation}
\end{eqnarray}
where, for a matrix-valued function of the chemical potential, we define $\widetilde{M}(\mu)
=\left[C^{-1}\gamma^0M(-\mu)\gamma^0C\right]^T$.  It is simple to check that 
$\overline{\overline{M}}(\mu)=\widetilde{\widetilde{M}}(\mu)=M(\mu)$ and that
$\widetilde{\overline{M}}(\mu)=\overline{\widetilde{M}}(\mu)$.  With these definitions, the matrix of
real-time 2-point functions for a fermion in a CP-invariant theory can be expressed, after a spatial Fourier
transformation, as
\begin{equation}
{\cal S}^{(ab)}(t,t';\bbox{k})=
\pmatrix{{\cal H}(t,t';\bbox{k})&\widetilde{\overline{\cal H}}(t,t';\bbox{k})\cr
{\cal H}(t,t';\bbox{k})&\widetilde{\overline{\cal H}}(t,t';\bbox{k})\cr}\theta(t-t')+
\pmatrix{\widetilde{\cal H}(t',t;\bbox{k})&\widetilde{\cal H}(t',t;\bbox{k})\cr
\overline{\cal H}(t',t;\bbox{k})&\overline{\cal H}(t',t;\bbox{k})\cr}\theta(t'-t),
\label{structure1}
\end{equation}
with
\begin{equation}
{\cal H}(t,t';\bbox{k})=\int d^3x\ e^{-i\bbox{k}\cdot\bbox{x}}{\cal H}(\bbox{x},t;\bbox{0},t')\,.
\end{equation}
Equivalently, defining ${\cal H}^{(1)}(t,t';\bbox{k})={\cal H}(t,t';\bbox{k})$ and ${\cal H}^{(2)}(t,t';\bbox{k})=
\widetilde{\overline{\cal H}}(t,t';\bbox{k})$, we can write
\begin{equation}
{\cal S}^{(ab)}(t,t')={\cal H}^{(b)}(t,t';\bbox{k})\theta(t-t')
+\widetilde{\cal H}^{(a)}(t',t;\bbox{k})\theta(t'-t)\,.
\label{structure2}
\end{equation}
We shall demand of our perturbative propagators that they have the structure shown
here (as, indeed, do the propagators of standard perturbation theory).  This does not mean
that our resummation can be applied only in the context of a CP-invariant theory;  it does 
mean, though, that any CP-violating effects will not be resummed.  It is worth pointing out
that a relation similar to (\ref{cprelation}) can be obtained by assuming C invariance rather
than CP invariance.  This would be equally usable, but phenomenologically a little more
restrictive.

The structure expressed by (\ref{structure1}) or (\ref{structure2}) implies two symmetries that
will be useful to us.  They are
\begin{equation}
\widetilde{\cal S}^{(ab)}(t,t';\bbox{k})={\cal S}^{(ba)}(t',t;\bbox{k})
\label{tildesymmetry}
\end{equation}
and
\begin{equation}
\pmatrix{
\overline{\cal S}^{(11)}(t,t';\bbox{k})&\overline{\cal S}^{(12)}(t,t';\bbox{k})\cr
\overline{\cal S}^{(21)}(t,t';\bbox{k})&\overline{\cal S}^{(22)}(t,t';\bbox{k})}
=\pmatrix{
{\cal S}^{(22)}(t',t;\bbox{k})&{\cal S}^{(12)}(t',t;\bbox{k})
\vphantom{\overline{\cal S}^{(22)}}\cr
{\cal S}^{(21)}(t',t;\bbox{k})&{\cal S}^{(11)}(t',t;\bbox{k})
\vphantom{\overline{\cal S}^{(22)}}}\,.
\label{barsymmetry}
\end{equation}
The first of these generalizes to the full $3\times 3$ matrix of real- and imaginary-time
2-point functions, but the second makes sense only for the real-time functions.

Finally, we shall need two pieces of information concerning the values of these functions
at equal times.  The functions ${\cal S}^{(12)}$ and ${\cal S}^{(21)}$ have unique values
at $t=t'$, which implies
\begin{equation}
\overline{\cal H}(t,t;\bbox{k})={\cal H}(t,t;\bbox{k})\,.
\label{equaltimebar}
\end{equation}
On the other hand, the time-ordered function ${\cal S}^{(11)}$ has a discontinuity at $t=t'$,
which reproduces the equal-time anticommutator
\begin{equation}
\lim_{t'\to t-0}{\cal S}^{(11)}_{\alpha\beta}(\bbox{x},t;\bbox{x}',t')
-\lim_{t'\to t+0}{\cal S}^{(11)}_{\alpha\beta}(\bbox{x},t;\bbox{x}',t')
=\langle\{\psi_\alpha(\bbox{x},t),\overline{\psi}_\beta(\bbox{x}',t)\}\rangle_\mu
=\gamma^0_{\alpha\beta}\delta(\bbox{x}-\bbox{x}')\,,
\end{equation}
and this implies
\begin{equation}
{\cal H}(t,t;\bbox{k})-\widetilde{\cal H}(t,t;\bbox{k})=\gamma^0\,.
\label{equaltimetilde}
\end{equation}

\section{Construction of the quasiparticle action}

We wish to construct a lowest-order action
\begin{equation}
I_{\rm C0}(\psi) = I_{\rm C}^{(2)}(\psi)+\int d^4x\,\overline{\psi}_a{\cal M}_{ab}\psi_b\equiv \int d^4x\,
\overline{\psi}_a{\cal D}_{ab}\psi_b
\label{unpertaction}
\end{equation}
that will serve as a starting point for our partially resummed perturbation theory.  As before,
$I_{\rm C}^{(2)}$ is the quadratic part of the original closed-time-path action, while the term involving
${\cal M}_{ab}$ is a counterterm which will be subtracted from the interaction part, so as to
leave the whole theory unchanged.  Specifying the form of ${\cal M}_{ab}$ is equivalent to
specifying the resulting differential operator ${\cal D}_{ab}$.  To begin, we construct the real-time
components of ${\cal D}_{ab}$, with $a,b=1,2$.  The unperturbed propagator matrix $S^{(ab)}(t,t')$
is a solution of the equations
\begin{equation}
{\cal D}_{ac}(t,\partial_t)S^{(cb)}(t,t')=S^{(ac)}(t,t'){\cal D}_{cb}(t', -\overleftarrow{{\partial}_{t'}})
=i\delta_{ab}\delta(t-t')\,.
\label{dseqdelta}
\end{equation}
(For economy of notation, we shall usually not indicate explicitly the dependence of these
quantities on $\bbox{k}$ and $\mu$.)  The form that might usefully be chosen for ${\cal D}_{ab}$
is constrained to a considerable extent by the requirement that this equation have solutions for
$S^{(ab)}$ which have the structure exhibited in (\ref{structure1}) and (\ref{structure2}) for the full
2-point functions and inherit the various properties that we discussed in Section III.  Observe first
that the $\delta(t-t')$ on the right of (\ref{dseqdelta}) arises from differentiating $\theta(t-t')$ and
$\theta(t'-t)$.  We can ensure that only these $\delta$ functions will appear by restricting ${\cal D}$
to the form
\begin{equation}
{\cal D}=\pmatrix{i\gamma^0\partial_t&0\cr 0&-i\gamma^0\partial_t}+\cdots\ ,
\end{equation}
where the ellipsis indicates terms without time derivatives.  The coefficients $\pm i\gamma_0$ are
determined by the boundary condition (\ref{equaltimetilde}).  In principle, an ansatz using more time
derivatives (along with further boundary conditions to eliminate unwanted $\delta$ functions and their
derivatives) might be possible, but we have not found such a generalization tractable.  Next, if the solution of
(\ref{dseqdelta}) is to have the symmetry expressed by (\ref{tildesymmetry}), then ${\cal D}$ must
satisfy
\begin{equation}
\widetilde{\cal D}_{ab}(t,\partial_t)={\cal D}_{ba}(t,-\partial_t)\,,
\end{equation}
which ensures that $I_{\rm C 0}$ is CP invariant.  Similarly, the symmetry expressed by (\ref{barsymmetry})
implies
\begin{equation}
\pmatrix{\overline{\cal D}_{11}(t,\partial_t)&\overline{\cal D}_{12}(t,\partial_t)\cr
\overline{\cal D}_{21}(t,\partial_t)&\overline{\cal D}_{21}(t,\partial_t)}
=\pmatrix{ -{\cal D}_{22}(t,-\partial_t)&-{\cal D}_{12}(t,-\partial_t)\cr
-{\cal D}_{21}(t,-\partial_t)&-{\cal D}_{11}(t,-\partial_t)}\,.
\end{equation}
Finally, as for a scalar field, causality requires
\begin{equation}
{\cal D}_{11}(t,\partial_t)+{\cal D}_{12}(t,\partial_t)+{\cal D}_{21}(t,\partial_t)+{\cal D}_{22}(t,\partial_t)=0\,.
\end{equation}
The net effect of these considerations is that ${\cal D}$ can be written as
\begin{equation}
{\cal D}(t,\partial_t)=\pmatrix{i\gamma^0\partial_t+{\cal D}_1(t)&i{\cal D}_2(t)\cr
i\widetilde{\cal D}_2(t)&-i\gamma^0\partial_t-\overline{\cal D}_1(t)}\,,
\label{diffoperator}
\end{equation}
where ${\cal D}_1(t)$ and ${\cal D}_2(t)$ are subject to the constraints
\begin{eqnarray}
\widetilde{\cal D}_1(t)&=&{\cal D}_1(t)\nonumber\\
\overline{\cal D}_2(t)&=&{\cal D}_2(t)\label{restrictions}\\
{\cal D}_1(t)+i{\cal D}_2(t)&=&\overline{\cal D}_1(t)-i\widetilde{\cal D}_2(t)\,.\nonumber
\end{eqnarray}
When ${\cal D}$ has this structure, and the propagator $S(t,t')$ is written in the form 
(\ref{structure1}) in terms of a function $H(t,t')$ that approximates ${\cal H}(t,t')$, then 
equations (\ref{dseqdelta}) for the propagator reduce to
\begin{eqnarray}
\left[i\gamma^0\partial_t+{\cal D}_1(t)+i{\cal D}_2(t)\right]H(t,t')&=&0\label{firstt}\\
\left[i\gamma^0\partial_t+{\cal D}_1(t)\right]\widetilde{H}(t',t)+i{\cal D}_2(t)\overline{H}(t',t)&=&0\,.
\label{secondt}
\end{eqnarray}

At this point, the most general procedure would be to expand ${\cal D}_i(t)$ in terms of
a complete basis of Dirac matrices
\begin{equation}
{\cal D}_i(t)=\sum_{p=1}^{16}d^p_i(t,\bbox{k})\Gamma_p\,,
\end{equation}
the thirty-two undetermined functions $d^p_i(t,\bbox{k})$ being the analogues of the functions
$\alpha_k(t)$, $\beta_k(t)$ and $\gamma_k(t)$ that appeared in the scalar theory.  This general
problem is one that we have not found tractable.  To simplify matters, we use instead the smallest
subset of the Dirac algebra that closes under multiplication and under $\overline{\vphantom{M}\ \ }$ and 
$\widetilde{\ }$ conjugation, and that includes the matrices $\gamma^0$ and $\bbox{\gamma}\cdot
\bbox{k}$ that appear in the free theory.  For the operator ${\cal D}$, a convenient basis is
$\{1, \gamma^0, \Gamma_+, \Gamma_-\}$, where
\begin{equation}
\Gamma_{\pm}=\frac{1}{2\vert\bbox{k}\vert}\left(1\pm\gamma^0\right)\bbox{\gamma}\cdot\bbox{k}
\end{equation}
and $1$ denotes the unit matrix.  The conjugates of these matrices are $\overline{\gamma^0}=
-\widetilde{\gamma^0}=\gamma^0$, $\overline{\Gamma}_{\pm}=\Gamma_{\mp}$ and
$\widetilde{\Gamma}_{\pm}=\Gamma_{\pm}$.  Our ansatz for ${\cal D}$ is then
\begin{eqnarray}
{\cal D}_1(t)&=&i\left[\lambda(t)-\nu(t)\right]\gamma^0+\left[\sigma(t)-\epsilon(t)\right]\Gamma_+
+\left[\sigma^{\sharp}(t)+\epsilon^*(t)\right]\Gamma_--\left[\tau(t)+i\eta(t)\right]\label{d1}\\
{\cal D}_2(t)&=&i\nu(t)\gamma^0+\epsilon(t)\Gamma_+-\epsilon^*(t)\Gamma_-+i\eta(t)\,.\label{d2}
\end{eqnarray}
Although we have not indicated it explicitly, the coefficients depend on $\bbox{k}$ and $\mu$ as
well as on $t$.  For complex functions of $\mu$, we define $f^{\sharp}(\mu)=f^*(-\mu)$, so that
$\widetilde{f}(\mu)=f(-\mu)=f^{*\sharp}(\mu)$.  This is the most general ansatz that satisfies the
restrictions (\ref{restrictions}) on ${\cal D}_i(t)$, provided that  
\begin{eqnarray}
\lambda^\sharp=\lambda,\quad \tau^\sharp=\tau,&& \nu^*=\nu, \quad \eta^*=\eta\nonumber\\
\lambda+\lambda^*&=&\nu+\nu^\sharp\nonumber\\
\sigma-\widetilde{\sigma}&=&\epsilon-\widetilde{\epsilon}\label{parrest}\\
\tau-\tau^*&=&-i(\eta-\eta^\sharp)\,.\nonumber
\end{eqnarray}

\section{Solution for the real-time quasiparticle propagators}

Having constructed the unperturbed action (\ref{unpertaction}) in terms of the differential operator
${\cal D}(t,\partial_t)$ given by (\ref{diffoperator}) together with the ans\"atze (\ref{d1}) and (\ref{d2}),
we require a formal solution to equations (\ref{firstt}) and (\ref{secondt}) for the matrix-valued function
$H(t,t';\bbox{k})$ from which the quasiparticle propagator $S^{(ab)}(t,t';\bbox{k})$ is to be constructed
via (\ref{structure2}).  It proves convenient to reorganize our basis of Dirac matrices into the set
$\{\gamma_{\pm}\equiv\frac{1}{2}(1\pm\gamma^0), \Gamma_{\pm}\}$, expanding $H(t,t')$ as
\begin{equation}
H(t,t')=\exp\left(-\int_{t'}^t\lambda(t'')dt''\right)\left[A(t,t')\gamma_++B(t,t')\Gamma_-+C(t,t')\Gamma_+
-D(t,t')\gamma_-\right]\,.\label{hansatz}
\end{equation}
Noting that (\ref{firstt}) governs the dependence of $H(t,t')$ on its first time argument, while (\ref{secondt})
refers to the second time argument, we introduce the notation $\dot{A}(t,t')\equiv\partial_tA(t,t')$ and
$\overcirc{A}(t,t')\equiv\partial_{t'}A(t,t')$.  With this notation, (\ref{firstt}) can be written as
\begin{eqnarray}
i\pmatrix{\dot{A}\cr\dot{B}}&=&\bbox{T}\pmatrix{A\cr B}\label{abeqn}\\
i\pmatrix{\dot{C}\cr\dot{D}}&=&\bbox{T}\pmatrix{C\cr D}\,, \label{cdeqn}
\end{eqnarray}
where
\begin{equation}
\bbox{T}=\pmatrix{\tau & \sigma\cr \sigma^{\sharp} &-\tau}\,.\label{tdef}
\end{equation}
For orientation, we note that, in the absence of the counterterm ${\cal M}_{ab}$, we would have $\tau=m$
and $\sigma=-\vert\bbox{k}\vert$.  (Equations somewhat analogous to these have been obtained, for example,
by Sahni \cite{sahni84} in the course of solving the Dirac equation in certain curved spacetimes.) The matrix 
$\bbox{T}$ has the generalized Hermiticity property $\bbox{T}^{\ddag}=\bbox{T}$, where the operation $^\ddag$
is defined by taking the transpose, and replacing each element by its $^\sharp$ conjugate. (This reduces to
the usual Hermitian conjugate when the chemical potential vanishes.)  If the two-component vectors $\varphi$
and $\chi$ are solutions to $i\dot{\varphi}=\bbox{T} \varphi$, then it is simple to see that the inner product 
$(\varphi,\chi)=\varphi^\ddag\chi$ is preserved by the time evolution.  The eigenvalues of $\bbox{T}(t)$ are 
$\pm\Omega(t)$, where the time-dependent frequency $\Omega=\sqrt{\tau^2+\sigma^\sharp \sigma}$
satisfies $\Omega^\sharp=\Omega$, and the corresponding normalized and mutually orthogonal 
eigenvectors are
\begin{equation}
u^{(+)}(t)=\frac{1}{\sqrt{2\Omega(\Omega-\tau)}}\pmatrix{\sigma\cr \Omega-\tau}\qquad\qquad
u^{(-)}(t)=\frac{1}{\sqrt{2\Omega(\Omega-\tau)}}\pmatrix{-(\Omega-\tau)\cr \sigma^{\sharp}}
\end{equation}
Given some fixed time $t_0$, one can formally write exact solutions
\begin{equation}
\phi^{(\pm)}_{t_0}(t)=\bbox{\cal T}\exp\left[-i\int_{t_0}^t\bbox{T}(t')dt'\right]u^{(\pm)}(t_0)\,,
\label{exactmodes}
\end{equation}
which are positive- and negative-frequency solutions
\begin{equation}
\phi^{(\pm)}_{t_0}(t)\approx \exp\left[\mp i\int_{t_0}^t\Omega(t')dt'\right]u^{(\pm)}(t_0)\,,
\end{equation}
at times near $t_0$. More generally, we may choose an orthonormal basis
\begin{equation}
\phi_1(t)=\pmatrix{f(t)\cr g(t)},\qquad\qquad \phi_2(t)=\pmatrix{-g^\sharp(t) \cr f^\sharp(t)}\,,
\end{equation}
with $f^\sharp(t)f(t)+g^\sharp(t)g(t)=1$.  It is readily verified that $\phi_2(t)$ is a solution if $\phi_1(t)$ is, and that if
$\phi_1(t)=\phi^{(+)}_{t_0}(t)$ for some $t_0$, then $\phi_2(t)= \phi^{(-)}_{t_0}(t)$.  Thus, the solution to
(\ref{abeqn}) and (\ref{cdeqn}) can be written as
\begin{eqnarray}
\pmatrix{A(t,t')\cr B(t,t')}&=&P_1(t')\pmatrix{f(t)\cr g(t)}+P_2(t')\pmatrix{-g^\sharp(t)\cr f^\sharp(t)}\label{absoln}\\
\pmatrix{C(t,t')\cr D(t,t')}&=&Q_1(t')\pmatrix{f(t)\cr g(t)}+Q_2(t')\pmatrix{-g^\sharp(t)\cr f^\sharp(t)}\label{cdsoln}\,,
\end{eqnarray}
where $P_i(t')$ and $Q_i(t')$ are to be determined by solving (\ref{secondt}) and applying suitable boundary
conditions.

To solve (\ref{secondt}), it is helpful to define
\begin{eqnarray}
W(t,t')&=&B^\sharp(t,t')-C(t,t')\nonumber\\
X(t,t')&=&D^\sharp(t,t')+A(t,t')\nonumber\\
Y(t,t')&=&B^\sharp(t,t')+C(t,t')\label{warelation}\\
Z(t,t')&=&D^\sharp(t,t')-A(t,t')\,.
\end{eqnarray}
In terms of these functions, (\ref{secondt}) is
\begin{eqnarray}
i\pmatrix{\overcirc{W}\cr\overcirc{X}}&=&\bbox{T}\pmatrix{W\cr X}\label{wxeqn}\\
i\pmatrix{\overcirc{Y}\cr\overcirc{Z}}&=&-i(\lambda+\lambda^*)\pmatrix{Y\cr Z} + \bbox{T'}\pmatrix{Y\cr Z}
+\bbox{E}\pmatrix{W\cr X}\label{yzeqn}\,.
\end{eqnarray}
The matrix $\bbox{T}$ is the same as the one defined in (\ref{tdef}), while
\begin{eqnarray}
\bbox{T'}&=&\pmatrix{\tau^* & \sigma+\epsilon^{*\sharp}-\epsilon\cr \sigma^\sharp + \epsilon^*-\epsilon^\sharp
& -\tau^*}\\
\bbox{E}&=&\pmatrix{i(\nu-\nu^\sharp-\eta-\eta^\sharp) & \epsilon^{*\sharp}+\epsilon\cr
-(\epsilon^*+\epsilon^\sharp) & i(\nu-\nu^\sharp+\eta+\eta^\sharp)}\,.
\end{eqnarray}
The appearance of a new matrix $\bbox{T'}$ (which also satisfies $\bbox{T'}^\ddag=\bbox{T'}$) and
a new damping constant $\lambda+\lambda^*$ in (\ref{yzeqn}) reflects the fact (previously noted
in \cite{lawrie97}) that particles and antiparticles acquire different thermal masses and widths in
the presence of a nonzero chemical potential.  The appearance of the inhomogeneous term
proportional to $\bbox{E}$ in (\ref{yzeqn}) implies that the time dependence of the propagators
is not exhausted by that of the single-particle mode functions, and we hope to interpret the
additional time dependence in terms of time-dependent occupation numbers for the single-particle
modes.  

Evidently, the solution of (\ref{yzeqn}) will require the introduction of a new set
of antiparticle mode functions, which make matters rather complicated.  In the remainder of this 
paper, we shall simplify our calculations by specializing to the case of zero chemical potential.
In that case, $^\sharp$ conjugation reduces to complex conjugation, while the restrictions
on the counterterm parameters noted in (\ref{parrest}) imply that $\tau$, $\lambda$ and $\eta$ are
real, and that $\nu=\lambda$.  We now have $\bbox{T'}=\bbox{T}=\bbox{T}^\dag$ and
\begin{equation}
\bbox{E}=\pmatrix{-2i\eta & 2\epsilon\cr -2\epsilon^* & 2i\eta}=-\bbox{E}^\dag\,.
\end{equation}
We look for a solution to (\ref{wxeqn}) and (\ref{yzeqn}) in the form
\begin{eqnarray}
\pmatrix{W(t,t')\cr X(t,t')}&=&R_1(t)\pmatrix{f(t')\cr g(t')}+R_2(t)\pmatrix{-g^*(t')\cr f^*(t')}\\
\pmatrix{Y(t,t')\cr Z(t,t')}&=&S_1(t,t')\pmatrix{f(t')\cr g(t')}
+S_2(t,t')\pmatrix{-g^*(t')\cr f^*(t')}\,,
\end{eqnarray}
and find that they are satisfied if 
\begin{eqnarray}
S_1(t,t')&=&S_1(t)L(t') + R_1(t)K_1(t')+R_2(t)K_2(t')\\
S_2(t,t')&=&S_2(t)L(t')+R_1(t)K_2^*(t')-R_2(t)K_1(t')\,,
\end{eqnarray}
where
\begin{eqnarray}
L(t')&=&\exp\left[-2\int_0^{t'}\lambda(t'')\,dt''\right]\,,\\
K_i(t')&=&L(t')\int_0^{t'}L^{-1}(t'')e_i(t'')\,dt''
\end{eqnarray}
and the $e_i(t)$ are related to the matrix elements $e_{ij}(t) = \phi_i^\dag(t)\bbox{E}(t)\phi_j(t)$ by
\begin{eqnarray}
e_{11}&=&e_{22}^*=2\left[\epsilon f^*g-\epsilon^*fg^*-i\eta(t')(f^*f-g^*g)\right]\equiv ie_1\label{e1def}\\
e_{12}&=&-e_{21}^*=2\left[\epsilon f^*f^* +\epsilon^*g^*g^*+i\eta f^*g^*\right]\equiv ie_2\,.\label{e2def}
\end{eqnarray}

At this point, the solution of (\ref{firstt}) (expressed by (\ref{abeqn}) and (\ref{cdeqn})) has left us with
four undetermined functions $P_i(t')$ and $Q_i(t')$, while the solution of (\ref{secondt}) (expressed by
(\ref{wxeqn}) and (\ref{yzeqn})) produced another four undetermined functions $R_i(t)$ and $S_i(t)$.
However, the two solutions are related by (\ref{warelation}) and by comparing them, we determine
all eight functions up to constants of integration.  Moreover, the values that these constants can
take are constrained by the equal-time conditions (\ref{equaltimebar}) and (\ref{equaltimetilde}), which
express very general properties of the 2-point functions. In fact, it turns out that that the propagator is
determined up to two constants of integration, that we shall denote by $a_1$ (which is real) and $a_2$
(which is complex).  Our final result for $H(t,t')$ is expressed by the original ansatz (\ref{hansatz}) with
the time-dependent coefficients given by
\begin{eqnarray}
A(t,t')&=&\left[1-N(t')\right]f(t)f^*(t')+N(t')g^*(t)g(t')-\Delta(t')f(t)g(t')-\Delta^*(t')g^*(t)f^*(t')\label{aresult}\\
B(t,t')&=&\left[1-N(t')\right]g(t)f^*(t')-N(t')f^*(t)g(t')-\Delta(t')g(t)g(t')+\Delta^*(t')f^*(t)f^*(t')\label{bresult}\\
C(t,t')&=&\left[1-N(t')\right]f(t)g^*(t')-N(t')g^*(t)f(t')+\Delta(t')f(t)f(t')-\Delta^*(t')g^*(t)g^*(t')\label{cresult}\\
D(t,t')&=&\left[1-N(t')\right]g(t)g^*(t')+N(t')f^*(t)f(t')+\Delta(t')g(t)f(t')+\Delta^*(t')f^*(t)g^*(t')\,.\label{dresult}
\end{eqnarray}
The real function $N(t')$ and the complex function $\Delta(t')$ are given, in terms of the quantities
introduced above, by
\begin{eqnarray}
N(t')&=&\frac{1}{2}\left[1-K_1(t')-a_1L(t')\right]\\
\Delta(t')&=&\frac{1}{2}\left[K_2(t')+a_2L(t')\right]\,,
\end{eqnarray}
but it is more illuminating to observe that they obey the differential equations
\begin{eqnarray}
\partial_{t'}N_k(t')&=&-2\lambda_k(t')N_k(t')+\left[\lambda_k(t')-\frac{1}{2}e_{1k}(t')\right]\label{dndt}\\
\partial_{t'}\Delta_k(t')&=&-2\lambda_k(t')\Delta_k(t')+\frac{1}{2}e_{2k}(t')\,,\label{ddeltadt}
\end{eqnarray}
with initial conditions $N_k(0)=\frac{1}{2}(1-a_{1k})$ and $\Delta_k(0)=\frac{1}{2}a_{2k}$, in
which we have reinstated the dependence on spatial momentum $k$ that has been implicit
throughout.  These are the equations that we might hope to interpret as kinetic equations for
the occupation numbers of single-particle modes.  Let us, indeed, specialize to the case of
thermal equilibrium, and temporarily delete the counterterm ${\cal M}_{ab}$.  The mode
functions can be written as
\begin{equation}
\pmatrix{f_k(t)\cr g_k(t)}=\frac{e^{-i\Omega_kt}}{\sqrt{2\Omega_k(\Omega_k-m)}}
\pmatrix{-\vert\bbox{k}\vert\cr \Omega_k-m}\,,
\end{equation}
with $\Omega_k=\sqrt{k^2+m^2}$ and we have $K_{1k}=K_{2k}=0$ and $L_k=1$.  We then find
that that the time-ordered function $S^{(11)}(t,t',\bbox{k}) =H(t,t';\bbox{k})\theta(t-t')
+\widetilde{H}(t',t;\bbox{k})\theta(t'-t)$, with $H(t,t';\bbox{k})$ given by (\ref{hansatz}) agrees with the
corresponding function obtained in \cite{kobes85}, provided that we can identify $N_k=\frac{1}{2}(1-a_1)
=N^{\rm eq}_k$ and $\Delta_k=\frac{1}{2}a_2=0$, where $N^{\rm eq}_k=\left[\exp(\beta\Omega_k)+1\right]^{-1}$ is
the usual Fermi-Dirac distribution.  The other 2-point functions do not agree with those of \cite{kobes85},
because these authors made use of a different time path (which is legitimate in thermal equilibrium, but
not in the nonequilibrium situation considered here).  At this stage, these values of the constants of 
integration $a_1$ and $a_2$ are merely guesses that yield this agreement with \cite{kobes85}.  The actual
values that are required by our formalism are determined by computing imaginary-time correlators and
applying appropriate boundary conditions.  This computation is the subject of the following section, where
we shall find our guesses confirmed.

\section{Imaginary and mixed-time propagators}
In terms of the imaginary-time field operator $\psi(\bbox{x},\tau)=e^{\widehat{H}\tau}\psi(\bbox{x},0)
e^{-\widehat{H}\tau}$, the imaginary- and mixed-time 2-point functions are (for $a=1,2$)
\begin{eqnarray}
{\cal S}^{(33)}_{\alpha\beta}(\bbox{x},\tau;\bbox{x}',\tau')&=&\langle\psi_\alpha(\bbox{x},\tau)
\overline{\psi}_\beta(\bbox{x}',\tau')\rangle\theta(\tau-\tau')-\langle\overline{\psi}_\beta(\bbox{x}',\tau')
\psi_\alpha(\bbox{x},\tau)\rangle\theta(\tau'-\tau)\\
{\cal S}^{(a3)}_{\alpha\beta}(\bbox{x},t;\bbox{x}',\tau')&=&-\langle\overline{\psi}_{\beta}(\bbox{x}',\tau')
\psi_\alpha(\bbox{x},t)\rangle\\
{\cal S}^{(3a)}(\bbox{x},\tau;\bbox{x}',t')&=&\widetilde{\cal S}^{(a3)}(-\bbox{x}',t';-\bbox{x},\tau)\,.
\end{eqnarray}
Because the time-path ordering makes imaginary times later than real times, the functions
${\cal S}^{(13)}$ and ${\cal S}^{(23)}$ are identical.  With $\mu=0$, antiperiodicity of the path integration
variables, $\psi_3(\bbox{x},\beta)=-\psi_1(\bbox{x},0)$, and the fact that the field operator $\psi(\bbox{x},0)$
is unique supply the two boundary conditions
\begin{eqnarray}
{\cal S}^{(a3)}(t,\beta;\bbox{k})&=&-{\cal S}^{(a1)}(t,0;\bbox{k})\label{bc1}\\
{\cal S}^{(a3)}(t,0;\bbox{k})&=&{\cal S}^{(a2)}(t,0;\bbox{k})\label{bc2}
\end{eqnarray}
that we shall use to determine the constants of integration $a_1$ and $a_2$.
If the sources for real-time fields are set to zero, then the time path reduces to just its imaginary-time
segment, and antiperiodicity yields
\begin{equation}
{\cal S}^{(33)}(0,\tau';\bbox{k})=-{\cal S}^{(33)}(\beta,\tau';\bbox{k})\,.\label{bc3}
\end{equation}
Finally, uniqueness of $\psi(\bbox{x},0)$ also implies
\begin{equation}
{\cal S}^{(a3)}(0,\tau';\bbox{k})={\cal S}^{(33)}(0,\tau';\bbox{k})\label{bc4}
\end{equation}
and the two latter boundary conditions serve to fix constants of integration that arise in the calculation
of ${\cal S}^{(a3)}$ and ${\cal S}^{(33)}$.

In order to construct a tractable perturbation theory, we have insisted that the unperturbed action
(\ref{unpertaction}) be local in time.  With this restriction, there are no terms of the form 
$\overline{\psi}_a(t){\cal D}_{a3}\psi_3(\tau)$, ($a=1,2$) so we have
\begin{equation}
I_{\rm c0}(\psi)=\int d^3x\left[\sum_{a,b=1}^2\int dt\,\overline{\psi}_a{\cal D}_{ab}\psi_b +\int_0^\beta d\tau
\,\overline{\psi}_3{\cal D}_{33}\psi_3\right]\,.
\end{equation}
In the second term, which approximates the path integral representation of the initial density operator,
$\overline{\psi}_3{\cal D}_{33}\psi_3$ is $-i$ times the Euclidean version of the free part of (\ref{lagrangian}),
with $m=m(0)$, supplemented by a counterterm ${\cal M}_{33}$.  The form of ${\cal D}_{33}$ is determined
by the assumed CP invariance, $\widetilde{\cal D}_{33}(\partial_\tau)={\cal D}_{33}(-\partial_\tau)$, together
with the requirement that the quasiparticle energy $\Omega_0$ be equal to the $t\to 0$ limit of the energy that
appears in the real-time mode functions.  This yields
\begin{equation}
{\cal D}_{33}(\partial_\tau)=i\left[\gamma^0\partial_\tau - \sigma_{0}\Gamma_+
-\sigma_{0}^*\Gamma_-+\tau_{0}\right]\,,
\end{equation}
where $\sigma_{0}$ and $\tau_{0}$ are the $t\to 0$ limits of the parameters appearing in (\ref{d1}).  The new
propagators satisfy
\begin{eqnarray}
{\cal D}_{33}(\partial_\tau)S^{(33)}(\tau,\tau')&=&S^{(33)}(\tau,\tau'){\cal D}_{33}(-\overleftarrow{\partial}_{\tau'})
=i\delta(\tau-\tau')\\
\left[{\cal D}_{11}(\partial_t)+{\cal D}_{12}\right]S^{(13)}(t,\tau')&=&
S^{(13)}(t,\tau'){\cal D}_{33}(-\overleftarrow{\partial}_{\tau'})=0\,,
\end{eqnarray}
and these equations can be solved by the method explained in the previous section.  The imaginary-time 
propagator can be expressed as $S^{(33)}(\tau,\tau';\bbox{k})=H_3(\tau,\tau';\bbox{k})\theta(\tau-\tau')
+\widetilde{H}_3(\tau',\tau;\bbox{k})\theta(\tau'-\tau)$, and we look for solutions of the form
\begin{eqnarray}
H_3(\tau,\tau')&=&A_3(\tau,\tau')\gamma_++B_3(\tau,\tau')\Gamma_-+C_3(\tau,\tau')\Gamma_+
-D_3(\tau,\tau')\gamma_-\\
S^{(13)}(t,\tau')&=&\exp\left(-\int_0^t\lambda(t'')dt''\right)\left[A_{13}(t,\tau')\gamma_++B_{13}(t,\tau')\Gamma_-
+C_{13}(t,\tau')\Gamma_+-D_{13}(t,\tau')\gamma_-\right]\,.
\end{eqnarray}

Defining positive- and negative-frequency mode functions in imaginary time by
\begin{equation}
\pmatrix{f_{\rm I}(\tau)\cr g_{\rm I}(\tau)}=\frac{e^{-\Omega_0\tau}}{\sqrt{2\Omega_0(\Omega_0-\tau_0)}}
\pmatrix{\sigma_0\cr \Omega_0-\tau_0}\qquad
\pmatrix{-\bar{g}_{\rm I}(\tau)\cr \bar{f}_{\rm I}(\tau)}=\frac{e^{\Omega_0\tau}}{\sqrt{2\Omega_0(\Omega_0-\tau_0)}}
\pmatrix{-( \Omega_0-\tau_0)\cr \sigma_0^*}\,,
\end{equation}
the solutions subject to the boundary conditions (\ref{bc3}) and (\ref{bc4}) are given by
\begin{eqnarray}
A_3(\tau,\tau')&=&(1-N^{\rm eq})f_{\rm I}(\tau)\bar{f}_{\rm I}(\tau')+N^{\rm eq}\bar{g}_{\rm I}(\tau)g_{\rm I}(\tau')\\
B_3(\tau,\tau')&=&(1-N^{\rm eq})g_{\rm I}(\tau)\bar{f}_{\rm I}(\tau')-N^{\rm eq}\bar{f}_{\rm I}(\tau)g_{\rm I}(\tau')\\
C_3(\tau,\tau')&=&(1-N^{\rm eq})f_{\rm I}(\tau)\bar{g}_{\rm I}(\tau')-N^{\rm eq}\bar{g}_{\rm I}(\tau)f_{\rm I}(\tau')\\
D_3(\tau,\tau')&=&(1-N^{\rm eq})g_{\rm I}(\tau)\bar{g}_{\rm I}(\tau')+N^{\rm eq}\bar{f}_{\rm I}(\tau)f_{\rm I}(\tau')
\end{eqnarray}
and
\begin{eqnarray}
A_{13}(t,\tau')&=& -N^{\rm eq} f(t)\bar{f}_{\rm I}(\tau')-(1-N^{\rm eq})g^*(t){g}_{\rm I}(\tau')\\
B_{13}(t,\tau')&=& -N^{\rm eq} g(t)\bar{f}_{\rm I}(\tau')+(1-N^{\rm eq})f^*(t){g}_{\rm I}(\tau') \\
C_{13}(t,\tau')&=& -N^{\rm eq} f(t)\bar{g}_{\rm I}(\tau')+(1-N^{\rm eq})g^*(t){f}_{\rm I}(\tau') \\
D_{13}(t,\tau')&=& -N^{\rm eq} g(t)\bar{g}_{\rm I}(\tau')-(1-N^{\rm eq})f^*(t){f}_{\rm I}(\tau')\,,
\end{eqnarray}
with $N^{\rm eq}=\left[e^{\beta\Omega_0}+1\right]^{-1}$.  Finally, the boundary conditions (\ref{bc1}) and 
(\ref{bc2}) are satisfied provided, as promised, that $a_1=1-2N^{\rm eq}$ and $a_2=0$.

\section{Concrete realization: A simple model}

In the preceding sections, we have constructed a quasiparticle action, and the propagators that
correspond to it, in terms of several time- and momentum-dependent coefficients that so far are
undetermined.  In general terms, these coefficients are to be determined self-consistently by 
asking the counterterm ${\cal M}_{ab}$ to cancel some part of the higher-order contributions to 
the self energy, which arise within an interacting theory that we also left  unspecified.  Thus, the
full 2-point functions can be expressed through the Schwinger-Dyson equation
\begin{equation}
{\cal S}^{(ab)}(t,t';\bbox{k})=S^{(ab)}(t,t',\bbox{k})-i\int dt''dt'''S^{(ac)}(t,t'';\bbox{k})\Sigma_{cd}(t'',t''';\bbox{k})
{\cal S}^{(db)}(t''',t';\bbox{k})\,,\label{sdeqn}
\end{equation}
in terms of a self energy that has contributions both from the counterterm and from loop corrections
\begin{equation}
\Sigma_{ab}(t,t')={\cal M}_{ab}(t,\partial_t)\delta(t-t')+\Sigma^{\rm loop}_{ab}(t,t')\,.
\end{equation}
If the counterterm could be chosen so that $\Sigma_{ab}=0$, then the propagator $S^{(ab)}$ would be
the same as the full 2-point function ${\cal S}^{(ab)}$ and the perturbation series would be completely
resummed.  In practice, of course, we can achieve at best a partial resummation by cancelling the first
few terms in the expansion of $\Sigma^{\rm loop}$.  Moreover, since $\Sigma^{\rm loop}$ is non-local in
time, and may well have a more complicated spinor structure than the counterterm we have constructed,
it will not be possible to cancel even these terms exactly. We can effect a selective resummation by 
cancelling only part of $\Sigma^{\rm loop}$, but the choice of which part to cancel will depend on details
of a specific application and of the supplementary approximations that will inevitably be required.  Here,
we illustrate how the process can be made to yield sensible results by studying the simplest possible 
model, in which our fermion interacts with a real scalar field, the interaction being specified by
\begin{equation}
{\cal L}_{\rm int}=-g\overline{\psi}\psi\phi\,.
\end{equation}
We suppose that the $\phi$ particles have a mass $M$ that is greater than $2m$, so that the decay and
annihilation processes $\phi\leftrightarrow\psi+\overline{\psi}$ are kinematically allowed.  We
anticipate that these on-shell processes will give absorptive parts to the fermion self energies,
yielding a nonzero thermal width $\lambda$, and that the equation (\ref{dndt}), which gives the rate of 
change of the occupation numbers $N_k(t)$, will, within a suitable approximation, be recognisable as
a kinetic equation of the Boltzmann type.  

We study explicitly the one-loop, real-time self energy, corresponding to the emission and reabsorption
of a $\phi$ particle.  By setting ${\cal M}+\Sigma^{\rm 1-loop}\approx 0$, we obtain a complicated set
of constraints, which implicitly specify the functions $\lambda_k(t)$, $\epsilon_k(t)$, etc.  In fact, these
functions enter $\Sigma^{\rm 1-loop}$ through the mode functions $f_k(t)$ and $g_k(t)$ and the
auxiliary functions $N_k(t)$ and $\Delta_k(t)$, for which we have no concrete expressions in hand.  We know
only that they are solutions of (\ref{abeqn}), (\ref{dndt}) and (\ref{ddeltadt}).  In principle, we have a closed set
of equations that we might attempt to solve numerically.  To see more clearly what these equations imply,
however, we introduce some further approximations.  First, we will suppose that time evolution is sufficiently
slow for an adiabatic approximation to be reasonable.  Then the mode functions will be written as
\begin{equation}
\pmatrix{f_k(t)\cr g_k(t)}\approx\frac{1}{\sqrt{2\Omega_k(t)(\Omega_k(t)-\tau_k(t))}}\pmatrix{\sigma_k(t)\cr
\Omega_k(t)-\tau_k(t)}\exp\left[-i\int_0^t\,\Omega_k(t')dt'\right]\,.
\end{equation}
Further, when the fermions have a nonzero thermal width, the propagator will decay at large time separations,
very roughly as $e^{-\lambda\vert t-t'\vert}$.  Assuming that $\tau_k(t)$ and $\sigma_k(t)$ do not change too
much over a thermal lifetime $\lambda_k(t)^{-1}$, it will be reasonable to approximate the product $f(t)f^*(t')$
that appears in (\ref{aresult}) as
\begin{equation}
f(t)f^*(t')\approx \frac{\sigma_k(\bar{t}\,)\sigma_k^*(\bar{t}\,)\exp\left[-i\Omega_k(\bar{t}\,)(t-t')\right]}
{2\Omega_k(\bar{t}\,) \left[\Omega_k(\bar{t}\,)-\tau_k(\bar{t}\,)\right]}\,,
\end{equation}
where $\bar{t}=(t+t')/2$, with corresponding approximations for other products of mode functions.  Finally, since
$\Sigma^{\rm 1-loop}$ is proportional to $g^2$, so are the functions $\lambda_k(t)$, etc that appear in ${\cal M}$.
At the lowest order of perturbation theory, it is therefore reasonably consistent to set these functions to zero in
the propagators that we use in evaluating $\Sigma^{\rm 1-loop}$, and this is what we do.  In particular, we then
have $\sigma_k(\bar{t})=-\vert\bbox{k}\vert$ and $\tau_k(\bar{t})=m(\bar{t}\,)$.  Clearly, these approximations are valid,
at best, only for a weakly interacting system in a state close to thermal equilibrium.  It is therefore important to 
emphasise that our purpose in introducing them is to obtain simple analytical results that illustrate essential 
features of our formalism.  The formalism itself is by no means restricted to situations where these approximations 
are valid.  For different reasons, discussed below, we will set $\Delta(t')=0$.

At this level of approximation, the propagators used in evaluating $\Sigma^{\rm 1-loop}$ (and those that multiply
$\Sigma_{ab}$ in (\ref{sdeqn})) are essentially those of the equilibrium theory, except that we allow for 
time-dependent masses and occupation numbers. It is useful to introduce the projection operators   
\begin{eqnarray}
\Lambda(\bbox{k})&=&\Omega_k(\bar{t}\,)\gamma^0-\bbox{\gamma}\cdot\bbox{k}+m(\bar{t}\,)\label{Lambda}\\
\widetilde{\Lambda}(\bbox{k})&=&-\Omega_k(\bar{t}\,)\gamma^0-\bbox{\gamma}\cdot\bbox{k}+m(\bar{t}\,)
\label{Lambdatilde}\\
\Lambda_{\pm}(\bbox{k})&=&\pm\left(\frac{m^2(\bar{t}\,)-k^2}{\Omega_k(\bar{t}\,)}\right)
\gamma^0-\bbox{\gamma}\cdot\bbox{k}+m(\bar{t}\,)\,,
\end{eqnarray}
which have the properties
\begin{eqnarray}
\Lambda(\bbox{k})\widetilde{\Lambda}(-\bbox{k})\Lambda(\bbox{k}) &=& 
\Lambda(\bbox{k})\Lambda_-(\bbox{k})\Lambda(\bbox{k})=0\label{proj1}\\
\widetilde{\Lambda}(\bbox{k})\Lambda(-\bbox{k})\widetilde{\Lambda}(\bbox{k}) &=& 
\widetilde{\Lambda}(\bbox{k})\Lambda_+(\bbox{k})\widetilde{\Lambda}(\bbox{k})=0\,.
\end{eqnarray}
After a Fourier transform on the time difference $t-t'$, each of the propagators appearing in the second term of
(\ref{sdeqn}) can be written in the form
\begin{equation}
S^{(ab)}(\bbox{k},\omega;\bar{t}\,)\approx S^{(ab)}_1(\bbox{k},\omega;\bar{t}\,)\Lambda(\bbox{k})
+S^{(ab)}_2(\bbox{k},\omega;\bar{t}\,)\widetilde{\Lambda}(\bbox{k})\,,\label{slambda}
\end{equation}
where $S^{(ab)}_1(\bbox{k},\omega;\bar{t}\,)$ has poles at $\omega=\Omega_k(\bar{t}\,)\pm i\lambda_k(\bar{t}\,)$, while
$S^{(ab)}_2(\bbox{k},\omega;\bar{t}\,)$ has poles at $\omega=-\Omega_k(\bar{t}\,)\pm i\lambda_k(\bar{t}\,)$ and, within 
the approximations described above, $\lambda_k(\bar{t}\,)$ is to be regarded as infinitesimal. In particular, we 
shall make explicit use of $S^{(12)}(\bbox{k},\omega;\bar{t}\,)$, in which the poles combine to yield
\begin{equation}
S^{(12)}(\bbox{k},\omega;\bar{t}\,)\approx \frac{\pi}{\omega}\left[\omega\gamma^0-\bbox{\gamma}\cdot\bbox{k}
+m(\bar{t}\,)\right]\left[N_k(\bar{t}\,)\delta\left(\omega-\Omega_k(\bar{t}\,) \vphantom{\bar{M^M_M}} \right)
+\left(1-N_k(\bar{t}\,)\vphantom{\bar{M^M_M}} \right)\delta\left(\omega+\Omega_k(\bar{t}\,
\vphantom{\bar{M^M_M}} )\right)\right]
\end{equation}
and of the corresponding scalar propagator
\begin{equation}
g^{(12)}(\bbox{k},\omega;\bar{t}\,)\approx \frac{\pi}{\omega_k(\bar{t}\,)}
\left[n_k(\bar{t}\,)\delta\left(\omega-\omega_k(\bar{t}\,) \vphantom{\bar{M^M_M}} \right)
+\left(1+n_k(\bar{t}\,)\vphantom{\bar{M^M_M}} \right)\delta\left(\omega+\omega_k(\bar{t}\,)
\vphantom{\bar{M^M_M}}\right)\right]\,,
\end{equation}
with $\omega_k(\bar{t}\,)=\sqrt{k^2+M^2(\bar{t}\,)}$.

We now consider how our vaguely-stated criterion ${\cal M}+\Sigma^{\rm loop} \approx 0$ can be implemented
in practice, to yield a selective resummation of the perturbation series.  To be clear, let us first reiterate the
r\^{o}le of the simplifying approximations introduced above.  In principle,   the functions $\tau_k(t)$, $\sigma_k(t)$,
$\lambda_k(t)\ \ldots$ are to be determined self-consistently by solving a set of equations of the form ${\cal M}
(\tau,\sigma,\lambda,\ldots) \approx -\Sigma^{\rm loop}(\tau,\sigma,\lambda,\ldots)$.  However, for the purposes
of discovering how $\approx$ might sensibly be interpreted and eventually of making contact with kinetic theory,
we plan instead to study the simplified set of equations ${\cal M}(\tau,\sigma,\lambda,\ldots)\approx 
-\Sigma^{\rm loop}(m,-\vert\bbox{k}\vert,0,\ldots)$, for which we can find closed-form expressions.  Of the
functions that we wish to determine, $\tau$ and $\sigma$ clearly encode the dispersion relation for the
quasiparticles whose mode functions are given by (\ref{exactmodes}).  These functions appear only in ${\cal D}_1$
(for which our ansatz was given in (\ref{d1})) and thus in ${\cal M}_{11}$ and ${\cal M}_{22}$.  By arranging for them
to cancel appropriate parts of $\Sigma^{\rm loop}_{11}$ and $\Sigma^{\rm loop}_{22}$, we can endow our
quasiparticles with dispersion relations that approximate those of the true elementary excitations of the
nonequilibrium state.  We do not do this explicitly, however, preferring to focus on dissipative aspects of the
problem, and especially on the evolution of occupation numbers.  In fact, for simplicity, we shall set
$\tau_k(t)=m(t)$ and $\sigma_k(t)=-\vert\bbox{k}\vert$ in ${\cal M}$ as well as in $\Sigma^{\rm loop}$.  The terms
proportional to $\Delta_k(t)$ in our propagators (\ref{aresult}) - (\ref{dresult}) are awkward and (as discussed in
the next section) difficult to interpret.  Within our present approximations, it is possible to eliminate these terms
in the following way.  We saw at the end of section VI that the constant $a_{2k}=\Delta_k(0)$ vanishes when
the initial state is one of thermal equilibrium.  It is therefore consistent to set $\Delta_k(t)=0$ at all times,
provided that the quantity $e_{2k}(t)$ vanishes in (\ref{ddeltadt}).  The expression given in (\ref{e2def}) does not
vanish in general, but with the approximations made here it can be made to do so by choosing $\epsilon_k(t)$
to be purely imaginary, say $\epsilon_k(t)=i\hat{\epsilon}_k(t)$, and by choosing 
$\eta_k(t)=-\tau_k(t)\hat{\epsilon}_k(t)/\sigma_k(t)=m(t)\hat{\epsilon}_k(t)/\vert\bbox{k}\vert$.

With these simplifications, the functions that remain to be determined are 
$\lambda_k(t)$ and $\hat{\epsilon}_k(t)$, which appear in the counterterm
\begin{equation}
{\cal M}_{12}(\bbox{k},t)=\frac{i}{2\Omega_k(t)}\left[ \left(\lambda_k(t)
+\frac{\Omega_k(t)}{\vert\bbox{k}\vert}\hat{\epsilon}_k(t)\right)
\Lambda(-\bbox{k})-\left(\lambda_k(t)-\frac{\Omega_k(t)}{\vert\bbox{k}\vert}\hat{\epsilon}_k(t)\right)
\widetilde{\Lambda}(-\bbox{k})\right]\,.\label{appm12}
\end{equation}
Ideally, we would like this to cancel the one-loop contribution
\begin{equation}
\Sigma^{\rm 1-loop}_{12}(\bbox{k},\omega;\bar{t}\,)
\approx -ig^2\int\frac{d\omega'\,d\omega''}{2\pi}\,\int\frac{d^3k'\,d^3k''}{(2\pi)^3}
\delta(\omega-\omega'-\omega'')\delta(\bbox{k}-\bbox{k}'-\bbox{k''})\,S^{(12)}(\bbox{k}',\omega';\bar{t}\,)
g^{(12)}(\bbox{k}'',\omega'';\bar{t}\,)\,,\label{appsig12}
\end{equation}
but ${\cal M}_{12}$, being derived from a counterterm that is local in time, has no dependence on $\omega$,
while $\Sigma^{\rm 1-loop}_{12}$ cannot be expressed in the form of (\ref{appm12}) as a linear combination of
$\Lambda(-\bbox{k})$ and $\widetilde{\Lambda}(-\bbox{k})$.  Clearly, we must be a little less ambitious.  First,
we shall attempt to effect the desired cancellation on shell:  that is, to cancel only 
$\Sigma^{\rm 1-loop}_{12}\left(\bbox{k},\pm\Omega_k(\bar{t}\,)\vphantom{\bar{M^M_M}}\right)$.  For
$\omega=\pm\Omega$, we find
\begin{eqnarray}
\Sigma^{\rm 1-loop}_{12}(\bbox{k},\omega;\bar{t}\,)\approx\frac{ig^2}{16\pi^2}\int d^3k'
\frac{\delta(\omega_p-\Omega-\Omega')}{\Omega'\omega_p}&&\left[ \left(-\Omega'\gamma^0
+\frac{\bbox{k}\cdot\bbox{k}'}{\vert\bbox{k}\vert^2}\bbox{\gamma}
\cdot\bbox{k}+m\right)n(1-N')\theta(\omega)\right.\nonumber\\
&&\qquad-\left.\left(\Omega'\gamma^0
+\frac{\bbox{k}\cdot\bbox{k}'}{\vert\bbox{k}\vert^2}\bbox{\gamma}\cdot\bbox{k}+m\right)
(1+n)N'\theta(-\omega)\right]\,.
\end{eqnarray}
Here, the kinematics is that of on-shell decay or pair annihilation, involving two fermions of momenta 
$\bbox{k}$ and $\bbox{k}'$, with energies $\Omega=\Omega_k(\bar{t}\,)$ and $\Omega'=\Omega_{k'}(\bar{t}\,)$,
and a scalar with momentum $\bbox{p}=\bbox{k}+\bbox{k}'$ and energy
$\omega_p=\sqrt{\vert\bbox{p}\vert^2 + M^2(\bar{t}\,)}$.  The fermion occupation numbers $N=N_k(\bar{t}\,)$,
and $N'=N_{k'}(\bar{t}\,)$ are those that we originally introduced in (\ref{aresult}) - (\ref{dresult}), while
$n=n_p(\bar{t}\,)$ is the corresponding quantity for the scalar.

When $\omega$ is close to $+\Omega_k$, the propagators given approximately by (\ref{slambda}) can be
further reduced by retaining only the term containing a pole, namely 
$S^{(ab)}\approx S^{(ab)}_1\Lambda(\bbox{k})$. The one-loop correction term in the Schwinger-Dyson equation
(\ref{sdeqn}) then involves the projection $\Lambda(\bbox{k})\left[{\cal M}_{12}+\Sigma^{\rm 1-loop}_{12}\right]
\Lambda(\bbox{k})$.  Using the properties (\ref{proj1}), we see that this can be made to vanish by expressing
$\Sigma^{\rm 1-loop}_{12}$ as a linear combination of $\Lambda(\bbox{k})$, $\widetilde{\Lambda}(-\bbox{k})$
and $\Lambda_-(\bbox{k})$ and requiring the coefficients of $\Lambda(\bbox{k})$ in $\Sigma^{\rm 1-loop}_{12}$
and ${\cal M}_{12}$ to cancel.  Similarly, when $\omega$ is close to $-\Omega_k$, we express 
$\Sigma^{\rm 1-loop}_{12}$ as a linear combination of $\widetilde\Lambda(-\bbox{k})$, ${\Lambda}(-\bbox{k})$
and $\Lambda_+(\bbox{k})$, and require the coefficients of $\widetilde\Lambda(-\bbox{k})$ to cancel.  In
this way, we obtain
\begin{eqnarray}
\lambda_k(\bar{t}\,)&=&\frac{g^2}{64\pi^2}(M^2-4m^2)\int d^3k'\frac{\delta(\omega_p-\Omega-\Omega')}
{\Omega\Omega'\omega_p}\left[n+N'\right]\label{lambdaresult}\\
\frac{\Omega_k(\bar{t}\,)}{\vert\bbox{k}\vert}\,\hat{\epsilon}_k(\bar{t}\,)   &=&
\frac{g^2}{64\pi^2}(M^2-4m^2)\int d^3k'\frac{\delta(\omega_p-\Omega-\Omega')}
{\Omega\Omega'\omega_p}\left[n(1-N')-(1+n)N'\right]\,.
\end{eqnarray}
 Here, we have made use of the kinematic identity $\Omega\Omega'-\bbox{k}\cdot\bbox{k}'-m^2=
\frac{1}{2}(M^2-4m^2)$.  Reassuringly, we have
arrived at a thermal quasiparticle width that is positive-definite when $M>2m$, so that the on-shell decay
and annihilation processes that give rise to it are kinematically allowed.  When $M<2m$, the thermal
width vanishes, because (\ref{appsig12}) contains products of $\delta$-functions that cannot be
satisfied simultaneously.  The expression (\ref{lambdaresult}) agrees with the equilibrium damping rate
calculated in \cite{boyanovsky99} (see also \cite{weldon83})for the same model.

Finally, we can evaluate the right hand side of equation (\ref{dndt}), which we hoped to interpret as
governing the evolution of time-dependent occupation numbers $N_k(t)$.  With the approximations
used in this section, we have $e_{1k}(t)=-2\Omega_k(t)\hat{\epsilon}_k(t)/\vert\bbox{k}\vert$, and the
equation becomes
\begin{equation}
\frac{d\,N_k(t)}{dt}=\frac{g^2}{32\pi^2}(M^2-4m^2)\int d^3k'\frac{\delta(\omega_p-\Omega-\Omega')}
{\Omega\Omega'\omega_p}\left[n(1-N)(1-N')-(1+n)NN'\right]\,.
\label{boltzmann}
\end{equation}
We recognise the standard form of a relativistic Boltzmann equation (restricted to a spatially
homogeneous system) in which the gain and loss terms have the correct statistical factors,
$n(1-N)(1-N')$ and $(1+n)NN'$ respectively, to represent a fermion of momentum $\bbox{k}$
being produced by the decay of a $\phi$ in the thermal bath or annihilating with an antifermion in
the bath.  Again, the `collision' integral would be replaced by 0 if $M<2m$.  Of course, our
results for $\lambda_k(t)$ and $dN_k(t)/dt$ were obtained only at the lowest non-trivial
order of perturbation theory.  At higher orders, we would expect non-zero answers in both
cases, arising from scattering processes that are allowed even for $M<2m$.

\section{Discussion}
We have described a selective resummation of the perturbation series for the nonequilibrium
2-point functions of spin-$\frac{1}{2}$ fermions.  The general philosophy of this resummation
is to describe the nonequilibrium state as nearly as possible in terms of its own quasiparticle
excitations.  The propagators for these excitations, unlike the free-particle propagators used
in standard perturbation theory ought, roughly speaking, to incorporate nonzero widths and
occupation numbers that evolve with time, reflecting the evolution of the nonequilibrium state.
The resummation is achieved through the use of a counterterm that transfers some
contributions of higher-order self energies into the lowest-order theory about which we perturb.
In a nonequilibrium situation, this is tractable only if the unperturbed action is local in time,
and this places strong constraints on what can be resummed in practice.  For example,
resummations somewhat similar in spirit to ours, but restricted to scalar theories in thermal
equilibrium, are described in \cite{banerjee91,parwani92,gleiser94} and applied to the
`warm inflation' scenario in \cite{berera98}.  In thermal equilibrium, the 2-point functions depend
only on $t-t'$, and after a Fourier transformation, one can (in principle) construct a counterterm
analogous to ${\cal M}$ that subtracts the whole frequency-dependent self energy at
whatever order of perturbation theory one has the energy to compute.  Generalizing this to
a nonequilibrium state would mean replacing (\ref{firstt}) and (\ref{secondt}) with 
integro-differential equations containing arbitrary non-local kernels in place of the functions
$\tau_k(t)$, $\lambda_k(t)$, etc. (or perhaps with infinite-order differential equations having
infinitely many time-dependent coefficients, all to be determined self-consistently) and this does
not seem to be a practical proposition.  For fermions, this may be particularly unfortunate, 
because the high-temperature plasma has, at least in some important cases, `hole' or `plasmino' 
excitations in addition to the particle and antiparticle poles that are present at $T=0$.  These have 
been known for some time from one-loop calculations \cite{lebellac96} and their properties have 
recently been explored in terms independent of perturbation theory by Weldon \cite{weldon00}.  
The counterterm we have constructed is linear in $\partial_t$ and cannot accommodate a
dispersion relation with these multiple branches (though a generalization that mimics them
might be possible).

Despite this inevitable deficiency, the resummation appears to make reasonable sense.  Of the
initially arbitrary parameters that we introduced in (\ref{d1}) and (\ref{d2}) to describe the quasiparticle
excitations, $\tau_k(t)$ and $\sigma_k(t)$ have clear interpretations in terms of the quasiparticle
dispersion relation, while $\lambda_k(t)$ is a thermal width which, after enough weak-coupling and
adiabatic approximations,  turns out to agree with the fermion damping rate calculated in equilibrium.
In the same approximation, the function $N_k(t)$ that appears in our resummed propagators
does indeed correspond to a quasiparticle occupation number, evolving according to a kinetic
equation of the Boltzmann type.  It is worth emphasizing that, while this kinetic equation
provides reassurance that our formalism has a sensible interpretation, its derivation is by no means
the purpose of the formalism presented here.  There are indeed, many routes to equations of this
kind.  One may, for example, investigate directly the time evolution of the expectation value of a
time-dependent number operator (see, e.g. \cite{boyanovsky96}).  Another route that has been pursued
extensively in connection with the calculation of transport coefficients of high-temperature plasmas
is to extract a transport equation for a Wigner density through truncation and gradient expansion of
the Schwinger-Dyson equations \cite{calzetta88,mrowczynski90,mrowczynski94,blaizot99}.  Attempts
to calculate transport coefficients directly from the Kubo formula of linear response theory reveal, on
the other hand, infrared singularities that require the resummation of large classes of diagrams
\cite{jeon93,jeon95} and this resummation turns out to be equivalent to solving a Boltzmann
equation \cite{jeon96}. (The relationship between these approaches is discussed in \cite{calzetta00}).
The transport equations that arise in these calculations go well beyond the simple one-loop
approximation exhibited in (\ref{boltzmann}), but they apply to systems very close to equilibrium.
Our own goal of constructing a resummed perturbation theory to describe the evolution of highly-excited
states that may be far from equilibrium is rather different.  The functions $N_k(t)$ that arise in the course
of solving for the resummed propagators are not necessarily equivalent to the Wigner distribution, and the
equation (\ref{dndt}) that defines them reduces to a Boltzmann equation only after approximations
that one may in general hope to avoid.

Finally, the investigation reported here indicates that the structure of nonequilibrium fermion propagators
is more complicated than might be expected from the equilibrium theory. Close to equilibrium, we found
that their spinor structure can be expressed in terms of the two projection operators (\ref{Lambda}) and
(\ref{Lambdatilde}), as can be deduced on general grounds within the equilibrium theory (see, e.g. 
\cite{weldon00}).  Away from equilibrium, this is no longer true.  In general, there are at least terms
proportional to $\gamma^0\bbox{\gamma}\cdot\bbox{k}$ and possibly other terms that we have not
succeeded in resumming.  To arrive at propagators that contain only $\Lambda(\bbox{k})$ and
$\widetilde{\Lambda}(\bbox{k})$, we need the coefficients $B(t,t')$ and $C(t,t')$ in (\ref{hansatz}) to
be equal.  This will be true of the solutions presented in (\ref{bresult}) and (\ref{cresult}) if the function 
$\Delta(t)$ vanishes, and if the products of mode functions $g(t)f^*(t')$ and $f(t)g^*(t')$ are equal.  Both
of these conditions would be automatic if we were to assume time-translation invariance, so that the
propagators depend only on $t-t'$, which is, of course, true in thermal equlibrium.  The equation
(\ref{ddeltadt}) satisfied by $\Delta(t)$ is superficially similar to (\ref{dndt}), but does not bear the same
interpretation as a kinetic equation.  In fact, the terms in the propagator that involve this function
are products (loosely speaking) of two positive-frequency or two negative-frequency mode functions
that have no counterpart in equilibrium.  In general, it seems that such terms should be present in
nonequilibrium self energies, but we have found no simple interpretation for them.

\acknowledgements

The final version of this work has benefitted from discussions with participants at the program on
Non-equilibrium Dynamics in Quantum Field Theory (1999) at the Institute for Nuclear Theory, 
University of Washington.  IDL would like to thank, in particular, Larry Yaffe and Francoise Guerin
for helpful comments and the INT for its hospitality.  DBM thanks the University of Leeds for
financial support through the award of a William Wright Smith scholarship.


\end{document}